\documentclass[reprint,aps,prd,twocolumn,amsmath,amssymb,nofootinbib]{revtex4-2}

\usepackage{graphicx}
\usepackage{hyperref}
\usepackage{booktabs}
\usepackage{xcolor}
\usepackage{tikz}
\usetikzlibrary{shapes.geometric, arrows.meta, positioning}

\hypersetup{
    colorlinks=true,
    linkcolor=blue,
    filecolor=magenta,
    urlcolor=cyan,
    citecolor=blue,
}

\begin{document}

\title{Detectability-Aware Screening of Post-Merger Gravitational-Wave Damping Anomalies in Real Detector Noise:\texorpdfstring{\\}{: }A Graph-Spectral Approach with Empirical Null Calibration}

\author{Ruslan Alaskarov}
\email{ruslanalas@gmail.com} % TODO: confirm contact details carried over from the earlier draft
\thanks{ORCID: \href{https://orcid.org/0009-0006-1030-3031}{0009-0006-1030-3031}. The author completed an M.Sc. in Physics at the Faculty of Physics, Lomonosov Moscow State University. The present work was conducted independently after graduation.}
\affiliation{Independent Researcher, Moscow, Russia} % TODO: update if necessary
\date{\today}

\begin{abstract}
The post-merger phase of binary neutron star (BNS) coalescences may encode information about matter at
extreme densities and could, in principle, be sensitive to non-standard energy-loss channels, including
scenarios motivated by dark-sector physics. Deviations in the effective damping time of post-merger
oscillations have been proposed as phenomenological proxies for such channels, but identifying short,
high-frequency transients in real, non-stationary detector noise remains difficult, and standard
statistical assumptions used to calibrate detection thresholds often do not hold in that regime.

We present a two-stage screening framework for a phenomenological damped-sinusoid post-merger proxy,
$h(t) = A e^{-t/\tau}\sin(2\pi f_{\rm peak} t + \phi)$, injected directly into real LIGO Livingston (L1)
background noise. Stage 1 builds a local time-frequency coherence graph from a Constant-Q representation
of each whitened candidate window and computes its leading eigenvalue as a detectability statistic,
motivated by the detectability-transition logic of spiked random matrices (the Baik--Ben Arous--P\'{e}ch\'{e}
transition) rather than a closed-form Gaussian-noise assumption. We validate the matched-filter SNR
normalization with a synthetic Gaussian-noise control, and separately validate the graph statistic with
power-comparison and tile-shuffling tests showing sensitivity to local time-frequency organization beyond
total integrated power, with the effect strongest for the shortest, most compact injected transients. Its
null distribution is calibrated empirically on real detector noise, since we find directly that real LIGO
noise does not satisfy the Gaussian assumptions a closed-form threshold would require. Stage 2 estimates
an effective damping time for Stage-1-passing candidates using an interpretable ridge/energy-decay fit
and a classical nonlinear least-squares fit, and compares each against a threshold calibrated on the
standard damping-time distribution ($\tau=5$--$40$~ms).

All thresholds and models are fixed on independent calibration and training data and evaluated once,
after the design was frozen, on a held-out real-noise segment untouched during development. The
least-squares estimator is the strongest Stage-2 discriminator (AUC $=0.943$); the ridge/energy-decay
estimator is a secondary, model-light, time-frequency-native alternative with competitive ranking-level
separation but lower recall at a matched low-false-positive operating point. A simple total-power
baseline is shown to be a real but substantially weaker, direction-sensitive discriminator.
\end{abstract}

\maketitle

\section{Introduction}

The direct detection of gravitational waves from the binary neutron star coalescence GW170817 initiated
multi-messenger constraints on the equation of state of dense matter \cite{Abbott2017PostMerger}. While
the inspiral phase constrains the macroscopic properties of cold neutron stars, the post-merger remnant
may encode information about matter at extreme temperatures and densities \cite{Bauswein2016HighDensityMatter}. Deviations
in the dominant post-merger frequency and effective damping time have been proposed as phenomenological
probes of the remnant's thermal and dynamical evolution; additional energy-loss channels, including
dark-sector degrees of freedom, could in principle modify this evolution and motivate searches for
anomalous damping. Post-merger emission is nonetheless difficult to detect with current interferometers:
the relevant frequencies are typically in the kilohertz range, where detector sensitivity is limited and
non-stationary instrumental artifacts complicate interpretation, and the true waveform depends on
dense-matter microphysics in ways that make exhaustive matched-filter searches computationally
challenging across broad classes of non-standard scenarios.

The purpose of this paper is methodological, not waveform-modeling. We do not attempt to model the full
post-merger dynamics of a binary neutron star remnant, whose structure, rotation, and quadrupole moment
evolve after merger and whose gravitational-wave emission can therefore include multiple modes, amplitude
modulation, and a time-dependent instantaneous frequency. Instead, we use a deliberately simple
damped-sinusoid injection family as a controlled proxy for short, high-frequency, post-merger-like
time-frequency excess, in order to isolate a narrower statistical question: whether an empirically
calibrated detectability pipeline can identify compact damping-time anomalies in real detector noise more
effectively than a power-only baseline. Section~\ref{sec:signal_model} states explicitly what this
simplification omits and where it is expected to fail; Sec.~\ref{sec:limitations} returns to this point
with a specific, mechanistic caveat about how an unmodeled frequency drift could bias the Stage-2 damping
estimator.

An earlier version of this project explored this problem using a frozen, pretrained audio-spectrogram
transformer as a generic time-frequency feature extractor, combined with a two-stage Mahalanobis
metric-learning framework. On review, that approach was found to have a manuscript/code mismatch (the
frozen backbone was pretrained on $\sim$10~s audio clips and applied to 100~ms windows, dominated by
padding), used an incorrect analytic detector-noise power spectral density (PSD), and did not adequately
account for close prior art applying pretrained audio transformers to gravitational-wave data
\cite{Chatterjee2025GWWhisper}. We abandoned that framing and rebuilt the analysis from the data-processing layer
up, adopting instead a detection-statistics approach with an explicit, testable connection to
random-matrix detectability theory and an empirically calibrated null distribution, motivated directly
by the finding (Sec.~\ref{sec:snr_validation}) that real detector noise does not satisfy the statistical
assumptions a closed-form Gaussian threshold would require.

In this work we treat post-merger anomaly screening as a two-stage detectability problem. Stage 1 asks
whether a candidate window's time-frequency content is organized in a way that is statistically
distinguishable from real detector noise, using a graph-spectral statistic inspired by (but not a literal
implementation of) the detectability-transition logic developed for spiked random matrices and,
separately, for community detection in the stochastic block model
\cite{BaikBenArousPeche2005,NadakuditiNewman2012,DecelleKrzakalaMooreZdeborova2011}. Stage 2 is applied
only to candidates that pass this gate and estimates an effective damping time, compared against a
threshold fixed on the standard ($\tau=5$--$40$~ms) manifold. Both stages are calibrated on real LIGO
noise and their thresholds are frozen before any evaluation on held-out data.

\section{Related Work}
\label{sec:related_work}

Unmodeled transient (``burst'') searches based on time-frequency excess power have a long history in
gravitational-wave data analysis, beginning with the excess-power statistic of
Anderson~\textit{et al.}~\cite{Anderson2001ExcessPower} and extended through coherent time-frequency
clustering methods such as Q-scan and coherent WaveBurst
\cite{Klimenko2008cWB,Chatterji2004QScan}, which were developed specifically to reduce sensitivity to
loud, non-Gaussian instrumental glitches relative to a simple power sum. The graph-based local-coherence
statistic used here for Stage~1 is a graph-spectral generalization in this same family, with an explicit,
empirically calibrated detectability threshold rather than a fixed analytic significance model.

Machine-learning approaches to compact-binary detection and parameter estimation on real LIGO data are
well established \cite{George2018DeepLearning}, and autoencoder-based anomaly detection has been applied
directly to LIGO glitch populations \cite{Laguarta2024GlitchAE}. Closest in spirit to this project's
originally abandoned framing, GW-Whisper~\cite{Chatterjee2025GWWhisper} applies a pretrained audio transformer,
fine-tuned with lightweight adapters, to gravitational-wave detection and glitch classification on real
O3b data, and reports that the model's native log-mel front end outperformed a GW-specific Q-scan variant
for glitch classification --- directly relevant to, and part of the motivation for, this project's
decision not to reintroduce a pretrained-audio-transformer component. Machine-learning parametrization of
post-merger waveforms via conditional variational autoencoders has also been explored
\cite{Whittaker2022PostMergerML}, and CNN-based post-merger detection/frequency-extraction methods have
been developed independently and concurrently with this work \cite{Weerasinghe2026PostMergerCNN}.

This project's contribution is not the general idea of time-frequency-structure-based transient
detection, nor the general idea of a pretrained-model feature extractor for gravitational-wave data ---
both have substantial prior art, cited above. The contribution is a specific combination: a
spiked-matrix/random-matrix-theory-motivated graph-spectral detectability statistic, calibrated
empirically on real LIGO noise and validated through power-comparison and tile-shuffling tests, together
with a matched-filter SNR convention separately checked using a synthetic Gaussian-noise control,
feeding an interpretable Stage-2 damping-time estimator whose thresholds are frozen before a single
held-out evaluation.

\section{Methodology}

\subsection{Phenomenological Signal Model}
\label{sec:signal_model}

We model the post-merger transient as a single-mode damped sinusoid,
\begin{equation}
    h(t) = A\, e^{-t/\tau} \sin(2\pi f_{\rm peak} t + \phi),
    \label{eq:signal_model}
\end{equation}
where $f_{\rm peak}$ is drawn uniformly from $[1500, 4000]$~Hz and $\tau$ is the effective damping time.
Equation~\ref{eq:signal_model} should be read as a controlled injection family, not as a physically
complete post-merger waveform model. It is a deliberately minimal phenomenological proxy, not a surrogate
for numerical-relativity post-merger waveforms, which exhibit multiple oscillation modes, a
time-dependent instantaneous frequency, and amplitude modulation tied to the evolving structure and
rotation of the remnant \cite{Bauswein2016HighDensityMatter,Whittaker2022PostMergerML}. The
constant-frequency assumption is used specifically to isolate the effects of damping time and
time-frequency localization under real-noise calibration, and is not intended to describe the early
dynamical evolution of a real merger remnant, where frequency drift, mode coupling, amplitude modulation,
and possible collapse to a black hole may all be important. Consequently, the performance reported below
is validated only for this simplified proxy family: in a realistically drifting signal, mismatch with the
constant-frequency least-squares model used in Stage 2 (Sec.~\ref{sec:stage2}) could bias the recovered
effective damping time toward shorter values, an effect not tested here and treated as a specific,
mechanistic limitation in Sec.~\ref{sec:limitations} rather than a settled question. Standard-damping
injections use $\tau \in \{5,10,20,40\}$~ms; anomalous injections use $\tau \in \{1,3\}$~ms.

\subsection{Data, PSD Estimation, and Whitening}
\label{sec:data_psd}

Real strain data are fetched from the LIGO Livingston (L1) detector via the Gravitational Wave Open
Science Center \cite{Abbott2021OpenData}, using its data-quality timeline to locate valid observing
segments rather than assuming an arbitrary offset from a reference GPS time is science-quality data ---
detector duty cycles are well below unity and hour-scale non-observing gaps are common. Four
non-overlapping segments are used, with fixed roles: PSD estimation, Stage-1 null calibration, Stage-2
training/calibration, and a final held-out evaluation segment untouched until Sec.~\ref{sec:final_eval}.
Table~\ref{tab:data_split} summarizes the role of each segment and exactly which downstream computations
each one contributes to, since the credibility of the held-out evaluation in Sec.~\ref{sec:results}
depends entirely on this split discipline being followed without exception.

\begin{table*}[htbp]
    \centering
    \caption{Data-split discipline. The Stage-2 calibration/validation halves are two non-overlapping
    subsegments of the single training segment referenced in the main text, split after PSD estimation
    (Sec.~\ref{sec:data_psd}) had already used the training segment as a whole; both halves therefore
    carry a ``yes'' for the PSD envelope, footnoted below. \texttt{test\_block} is excluded from every
    fitting step without exception, including the PSD envelope.}
    \label{tab:data_split}
    \begin{tabular}{@{}p{0.15\textwidth}p{0.08\textwidth}p{0.16\textwidth}p{0.08\textwidth}p{0.22\textwidth}p{0.14\textwidth}@{}}
        \toprule
        \textbf{Segment} & \textbf{Dur.} & \textbf{Role} & \textbf{PSD env.?} & \textbf{Threshold/model fit?} & \textbf{Final eval.?} \\
        \midrule
        \texttt{psd\_block} & 2048~s & PSD estimation & Yes & No & No \\
        \texttt{calib\_block} & 1024~s & Stage-1 null calibration & Yes & Yes (Stage-1 threshold) & No \\
        Training segment, calib.\ half$^{a}$ & 508~s & Stage-2 threshold / classifier fitting & Yes$^{a}$ & Yes (Stage-2 thresholds, power classifier) & No \\
        Training segment, valid.\ half$^{a}$ & 508~s & Held-out Stage-2 validation & Yes$^{a}$ & No (eval.\ only, no refitting) & No \\
        \texttt{test\_block} & 1024~s & Final evaluation & No & No & Yes, once, after freeze \\
        \bottomrule
    \end{tabular}
    \\[2pt]
    \raggedright\footnotesize $^{a}$Both halves were part of the single training segment used, as a
    whole, in the PSD envelope (Sec.~\ref{sec:data_psd}); the split into calibration/validation halves
    was performed afterward, specifically for the Stage-2 held-out validation described in
    Sec.~\ref{sec:final_eval}.
\end{table*}

The operational power spectral density (PSD) is estimated as an envelope (element-wise maximum) of
independent per-segment estimates across the PSD-estimation, calibration, and training segments (the
held-out evaluation segment is deliberately excluded from this envelope). Each per-segment estimate
combines a bias-corrected median (an accurate baseline away from spectral lines) with a
percentile-based intermittency detector: bins whose $99$th-percentile-across-Welch-segments value
substantially exceeds the typical ratio for stationary noise are flagged and inflated to a conservative
multiple of their own observed peak value. This combination was necessary because narrow instrumental
lines (violin-mode harmonics and their overtones) drift in frequency and amplitude over time scales
shorter than a single PSD-estimation segment, so a single-snapshot median estimate --- even a correctly
bias-corrected one --- can substantially underestimate a line's true peak amplitude in a different
segment.

Whitening is performed in the frequency domain, hard-band-limited to $[20, 6000]$~Hz (data outside this
range is never divided by a PSD estimate, since PSD estimates near DC and Nyquist are unreliable). Each
long data segment is whitened once, as a whole, and edge-trimmed before being sliced into $100$~ms
analysis windows; whitening a bare $100$~ms window in isolation was found to reintroduce a
circular-convolution edge artifact and is avoided throughout.

\subsection{Matched-Filter SNR and Injection}
\label{sec:snr_validation}

Injected signal-to-noise ratio is defined via the standard matched-filter integral,
\begin{equation}
    \rho^2 = 4 \int_0^\infty \frac{|\tilde h(f)|^2}{S_n(f)}\, df,
    \label{eq:mf_snr}
\end{equation}
discretized consistently with the whitening convention above. We validated this definition in two
stages, which we report because the distinction is central to this project's approach to null
calibration. First, we generated synthetic Gaussian noise colored to exactly match the assumed PSD and
FFT convention, and confirmed that recovered matched-filter SNR follows the expected
$\mathcal{N}(\rho_{\rm target}, 1)$ distribution essentially exactly (mean/median/std/robust-std all
within a few percent of the ideal values). Second, we repeated the identical recovery test on real L1
noise and found a substantially broader, non-Gaussian distribution (robust standard deviation an order of
magnitude larger than the ideal case, with pronounced asymmetric tails). Since the first test confirms
the SNR formula's normalization is correct, the second result is not a bug: it is direct evidence that
real detector noise, in this frequency band and window length, does not satisfy the stationary-Gaussian
assumption a closed-form detection threshold would require. This is the direct motivation for the
empirical, real-noise-calibrated approach used throughout Stage~1 and Stage~2.

\begin{figure*}[htbp]
    \centering
    \includegraphics[width=0.85\textwidth]{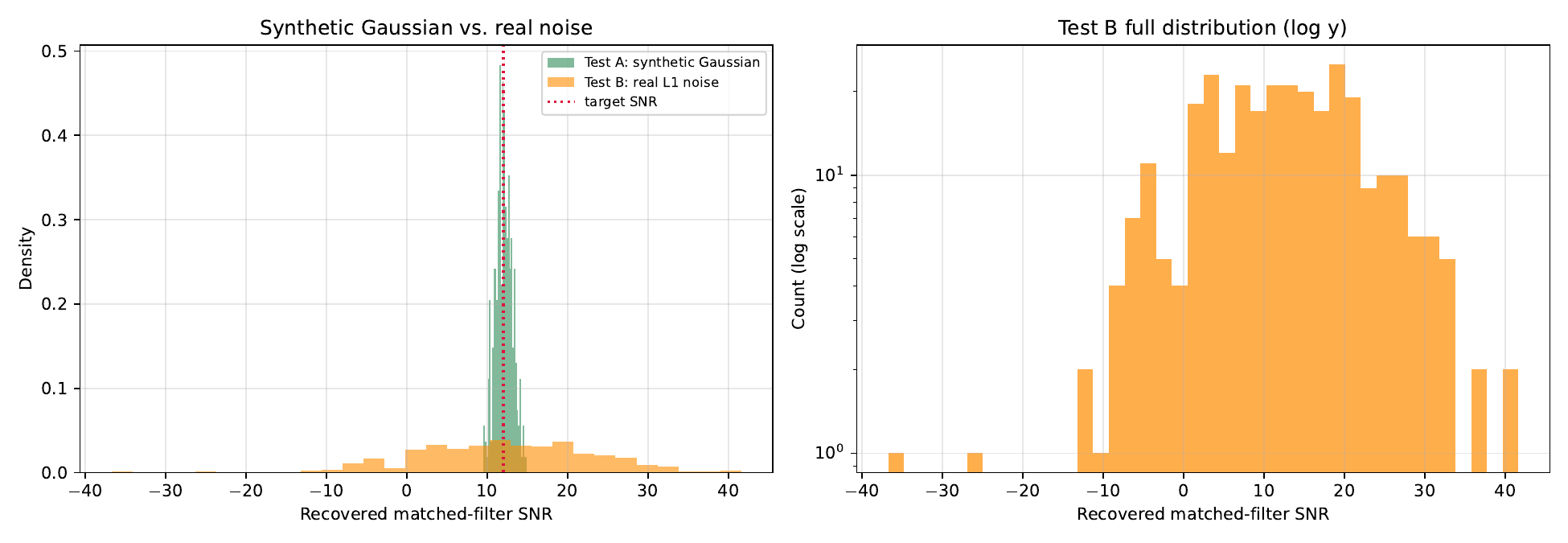}
    \caption{Matched-filter SNR recovery validation. \textbf{Left:} recovered SNR from a fixed injected
    target of $\rho=12$, comparing synthetic Gaussian noise built to exactly match the assumed PSD
    convention (tight peak at the target) against real L1 noise (broad, asymmetric spread). The synthetic
    case alone confirms the SNR formula's normalization is correct; the real-noise deviation is therefore
    direct evidence about the noise itself, not a normalization error. \textbf{Right:} the same real-noise
    distribution on a log scale, showing the extended, non-Gaussian tails.}
    \label{fig:snr_validation}
\end{figure*}

Injections are constructed by drawing a padded local noise segment from the target block's own noise
pool, injecting the scaled waveform at the window center, whitening the padded segment as a whole (for
the same edge-artifact reason as Sec.~\ref{sec:data_psd}), and cropping back to the $100$~ms analysis
window.

\subsection{Stage 1: Graph-Spectral Detectability}
\label{sec:stage1}

Each candidate window's whitened time series is converted to a Constant-Q Transform (CQT) magnitude map,
restricted to the post-merger band of interest ($1500$--$4000$~Hz), yielding a small set of frequency
bins by a modest number of time frames. Every tile's magnitude is normalized against an empirical null
built from many real noise-only windows: $Z = (\log(1+\text{CQT}) - \mathrm{median})/\mathrm{MAD}$,
floored at zero, so that only excess above the empirically observed noise level for that specific
time-frequency bin contributes.

\begin{figure*}[htbp]
    \centering
    \includegraphics[width=0.85\textwidth]{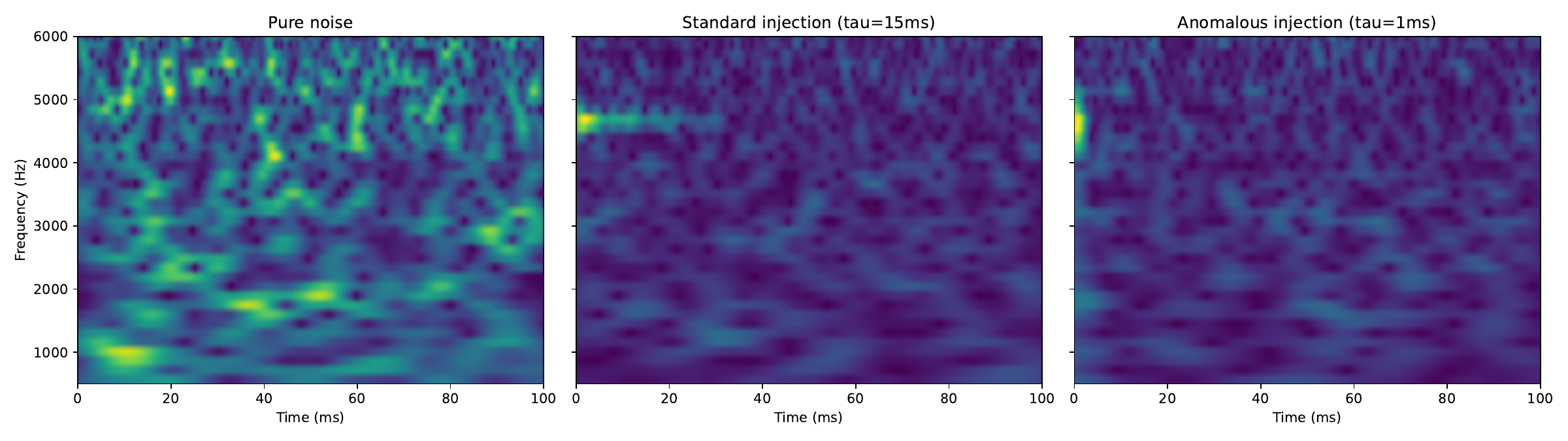}
    \caption{Example whitened Constant-Q representations shown over a wider visual frequency range for
    illustration; the Stage-1 graph statistic itself is computed only on the $1500$--$4000$~Hz band
    (Sec.~\ref{sec:stage1}). \textbf{Left:} pure real L1 noise. \textbf{Center:} a standard-damping
    injection ($\tau=15$~ms). \textbf{Right:} an anomalous-damping injection ($\tau=1$~ms), visibly more
    localized in time than the standard case at comparable amplitude.}
    \label{fig:cqt_examples}
\end{figure*}

We build a sparse graph over the normalized tile map, connecting each tile to others within a small local
time-frequency neighborhood, with edge weight equal to the product of the two tiles' (already
null-normalized) values --- a local ``bright-neighbors-of-bright-tiles'' coherence proxy. The Stage-1
detectability statistic is the leading eigenvalue of this graph's adjacency matrix. This construction is
motivated by, but is \emph{not} a literal implementation of, the detectability-transition logic developed
for spiked random matrices \cite{BaikBenArousPeche2005} and, in a related but distinct setting, for
community detection in the stochastic block model
\cite{DecelleKrzakalaMooreZdeborova2011,NadakuditiNewman2012}: a planted low-rank structure is invisible in
a matrix's leading eigenvalue until its strength crosses a threshold, a logic we exploit as a design
motivation and validate empirically rather than assume. We refer to this statistic throughout as
``Moore/BBP-inspired graph-spectral detectability,'' never as an implementation of stochastic-block-model
community detection.

We validated that this statistic is not merely a restatement of total time-frequency power in two
independent ways. First, across a mixed sample of noise-only and injected windows spanning several
damping times, total in-band power and the graph statistic showed a low pooled correlation ($r \approx
0.17$), and a controlled comparison at fixed matched-filter SNR (pooling across damping time) gave a
correlation consistent with zero ($r \approx 0.01$--$0.08$ across several SNR values) --- at matched
injected energy, total power carries essentially no information about the graph statistic. Second, and
more directly, we performed a tile-position shuffle test: keeping every CQT tile's value fixed (hence
total power exactly unchanged) while randomizing tile positions before rebuilding the graph produces
large, highly significant reductions in the leading eigenvalue for every category tested ($p<10^{-10}$,
Wilcoxon signed-rank), and the reduction is largest for the shortest, most compact injected transients
($\tau=1$~ms: $86$--$87\%$ reduction) and smallest, though still substantial, for pure noise ($54$--$64\%$
reduction). The correct, conservative statement supported by this evidence is that the graph statistic
captures local organization of time-frequency excess beyond total integrated power; it does not, by
itself, establish that the statistic measures physical post-merger coherence, ridge structure, or any
other specific physical mechanism, since compact transient localization and connected high-energy
clustering remain observationally indistinguishable candidate explanations. We adopt this conservative
interpretation throughout.

\begin{figure*}[htbp]
    \centering
    \includegraphics[width=0.85\textwidth]{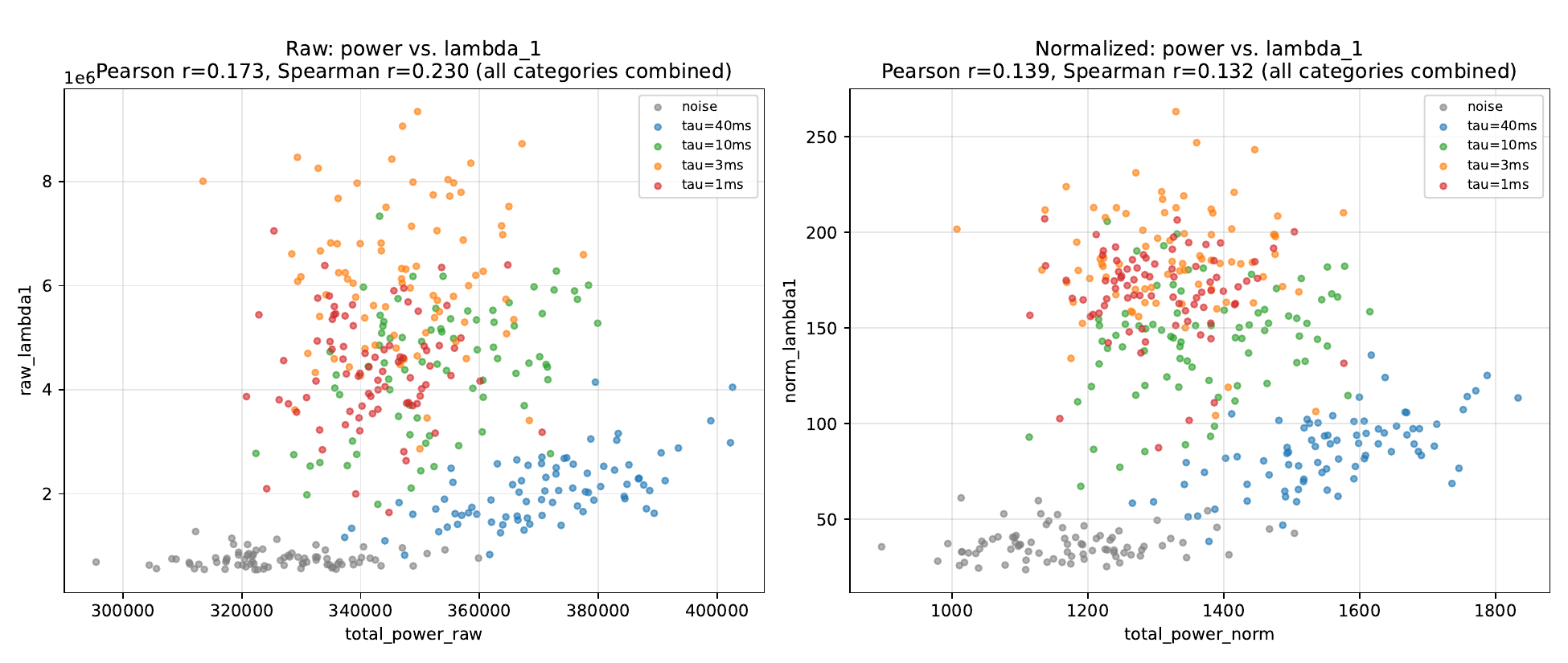}
    \caption{Total in-band power versus the graph-spectral statistic, raw (left) and null-normalized
    (right), across noise-only windows and four damping times. The low pooled correlation in both panels
    ($r=0.17$ raw, $r=0.14$ normalized) shows the two statistics are not redundant; note the categorical
    vertical separation by $\tau$ at fixed power, which the shuffle test (main text) further confirms is
    driven by spatial organization rather than total energy.}
    \label{fig:power_vs_lambda1}
\end{figure*}

The Stage-1 null distribution is calibrated empirically from several thousand real noise-only windows,
at a chosen false-alarm rate $\alpha=0.01$; the calibration sample size was chosen so that the expected
number of exceedances at this threshold is large enough for the resulting quantile estimate to be
reasonably stable, following the same reasoning used to size the calibration sample throughout this work.

\begin{figure*}[htbp]
    \centering
    \includegraphics[width=0.85\textwidth]{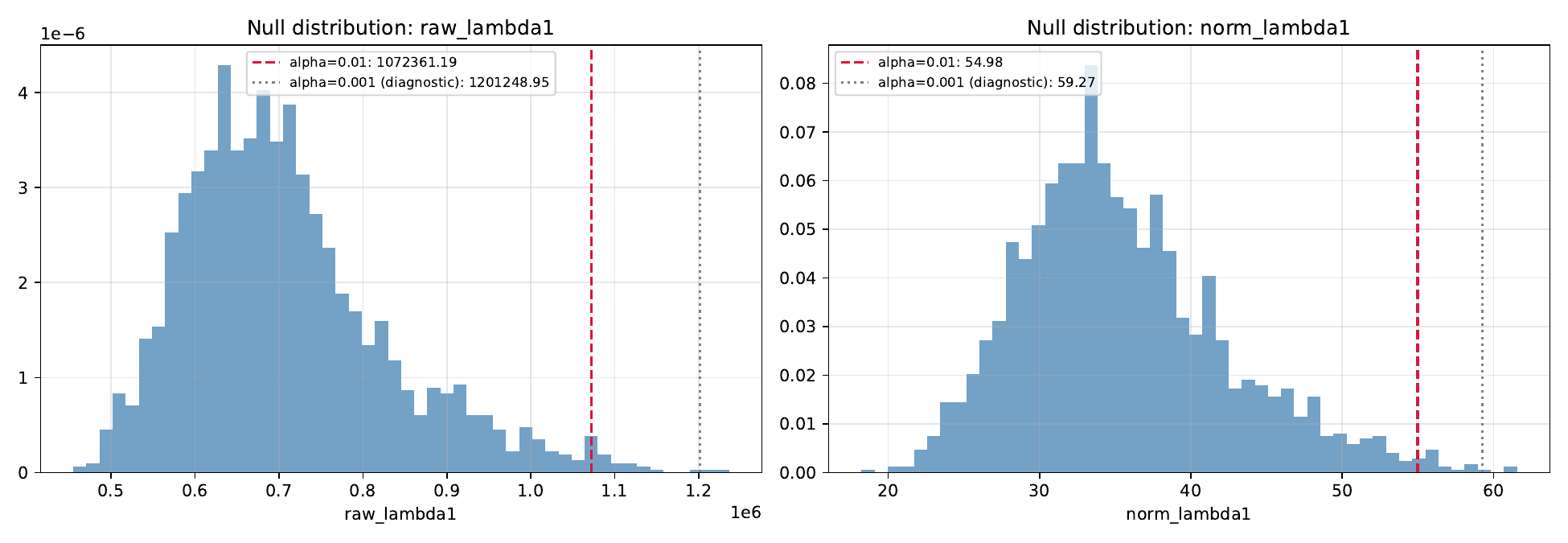}
    \caption{Empirical Stage-1 null distributions on real L1 noise, raw (left) and null-normalized
    (right) graph statistic, from an earlier exploratory calibration pass (threshold $\approx 54.98$ for
    the normalized statistic) shown here for illustration of the calibration procedure. The final frozen
    threshold used for the \texttt{test\_block} evaluation in Sec.~\ref{sec:results} is $55.6439$,
    calibrated with a larger sample under the same procedure; the two values agree to within the
    calibration sample's expected precision. The $\alpha=0.001$ line is shown for reference only; the
    calibration sample size used here is not adequate for a trustworthy $0.1\%$-tail estimate
    (Sec.~\ref{sec:stage1}) and $\alpha=0.001$ is never used for any reported result.}
    \label{fig:null_distributions}
\end{figure*}

\begin{figure*}[htbp]
    \centering
    \includegraphics[width=0.85\textwidth]{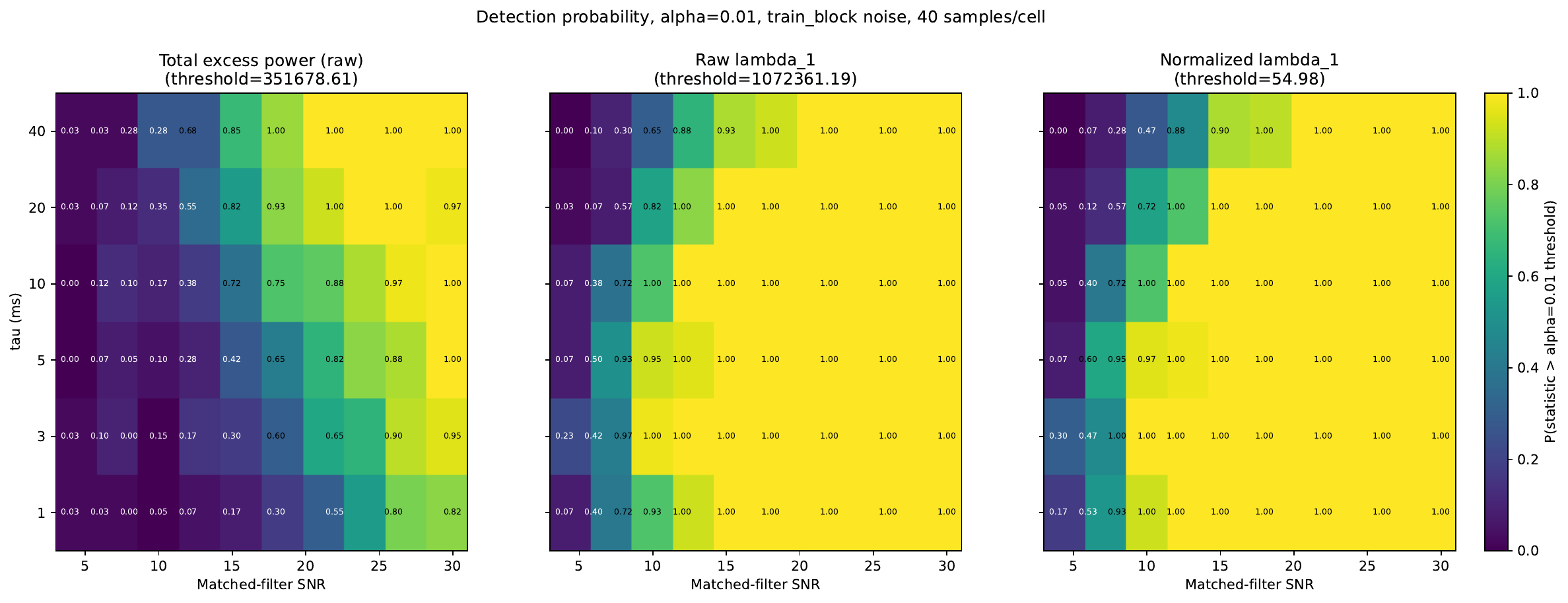}
    \caption{Detection probability at $\alpha=0.01$ over a $(\mathrm{SNR},\tau)$ grid, comparing total
    excess power (left), the raw graph statistic (center), and the null-normalized graph statistic
    (right). This is an earlier, exploratory training-data pass (threshold $\approx 54.98$ for the
    normalized statistic, before the calibration sample was enlarged for the final frozen threshold of
    $55.6439$ used in Sec.~\ref{sec:results}), shown to illustrate the qualitative sensitivity gap
    motivating Stage~1's design; it is distinct from, and should not be read as, the final held-out
    evaluation. At $\tau=1$~ms, total power never reaches $P_{\rm det}=0.9$ within the tested SNR range,
    while both graph-statistic variants reach it by SNR$\,\approx 8$--$10$ --- the same gap that
    motivates preferring a structure-sensitive statistic over a power baseline for Stage~1.}
    \label{fig:phase_diagram}
\end{figure*}

\subsection{Stage 2: Damping-Time Estimation}
\label{sec:stage2}

Stage 2 is evaluated only for candidates that pass the Stage-1 gate. We use two independent estimators of
an effective damping time $\hat\tau$:

\textit{Primary --- nonlinear least-squares fit.} A direct fit of Eq.~\ref{eq:signal_model} to the
whitened time-domain window via nonlinear least squares, using the in-band dominant frequency as an
initial guess for $f_{\rm peak}$.

\textit{Secondary --- ridge/energy-decay fit.} An interpretable, time-frequency-native alternative: the
in-band CQT power per time frame traces the signal envelope, and since $h(t)^2 \sim e^{-2t/\tau}$, a
linear fit of $\log(\text{power})$ against time (starting from the frame of peak power) recovers
$\hat\tau$ from the fitted slope. The fit window is truncated adaptively once frame power drops into a
noise-floor reference calibrated on real noise, rather than using a fixed-length window; an earlier fixed
$29$~ms window was found to produce severe overestimation of $\hat\tau$ for short-duration signals, since
most of the fitted region was flat noise floor rather than genuine decay for any $\tau$ shorter than a
few milliseconds.

Both estimators' anomaly thresholds are calibrated as the first percentile of their own $\hat\tau$
distribution on standard-damping ($\tau=5$--$40$~ms) Stage-1-passing candidates, drawn from an
independent calibration subset; a candidate is flagged anomalous if $\hat\tau$ falls below this
threshold. Neither estimator's internal parameters were tuned after this calibration was fixed.

\textit{Baseline --- total power.} As a baseline discriminator not aware of any temporal structure, we
fit a single-feature logistic regression on total in-band power alone, on the same calibration subset,
and calibrate its own operating threshold (at the $99$th percentile of predicted probabilities on the
calibration subset's standard-damping candidates) to target a comparable low false-positive rate to the
$\hat\tau$-based methods --- a fair operating-point comparison, since a classifier's default
decision-probability threshold does not, in general, correspond to any particular target false-positive
rate.

\subsection{Held-Out Validation and Final Evaluation}
\label{sec:final_eval}

Before any evaluation on the final held-out segment, we validated Stage~2 on an independent split within
the training data: all thresholds and the power baseline's classifier were fit on one half of the
training noise pool and evaluated, without refitting, on the other half. Results were consistent with the
in-sample development numbers to within sampling noise for every metric reported, indicating no
meaningful overfitting to the training data.

The complete design --- Stage-1 statistic and threshold, both Stage-2 estimators and their thresholds,
and the power baseline's classifier and operating threshold --- was then frozen and recorded in a
machine-readable configuration file. The held-out evaluation segment, untouched by any prior step in this
work, was loaded for the first time only after this freeze, used exclusively for the evaluation reported
in Sec.~\ref{sec:results}, and did not influence any threshold, model, or hyperparameter.

\section{Results}
\label{sec:results}

Table~\ref{tab:final_results} summarizes the final held-out evaluation. The least-squares estimator is
the strongest Stage-2 discriminator, both by ranking quality (AUC~$=0.943$) and, more importantly for a
practical screening threshold, by recall at the calibrated low-false-positive operating point
(TPR~$=0.879$ at FPR~$=0.012$). The ridge/energy-decay estimator provides strong ranking-level
separation (AUC~$=0.930$) but substantially lower recall at the same operating-point standard
(TPR~$=0.486$); we report both together throughout, since AUC alone does not convey this gap --- LSQ
gives the strongest low-FPR operating-point recall of the two. The calibrated power baseline contains
moderate, direction-sensitive ranking information (AUC~$=0.726$ in its correct orientation) but performs
poorly at the strict operating point relevant for screening (TPR~$=0.054$) --- a moderate ranking-quality
AUC is fully compatible with near-zero recall at a strict threshold when the underlying class
distributions overlap substantially in their bulk, and this comparison is, in our view, the most direct
evidence that a temporal-structure-aware estimator is necessary for this screening task.

\begin{figure}[htbp]
    \centering
    \includegraphics[width=\columnwidth]{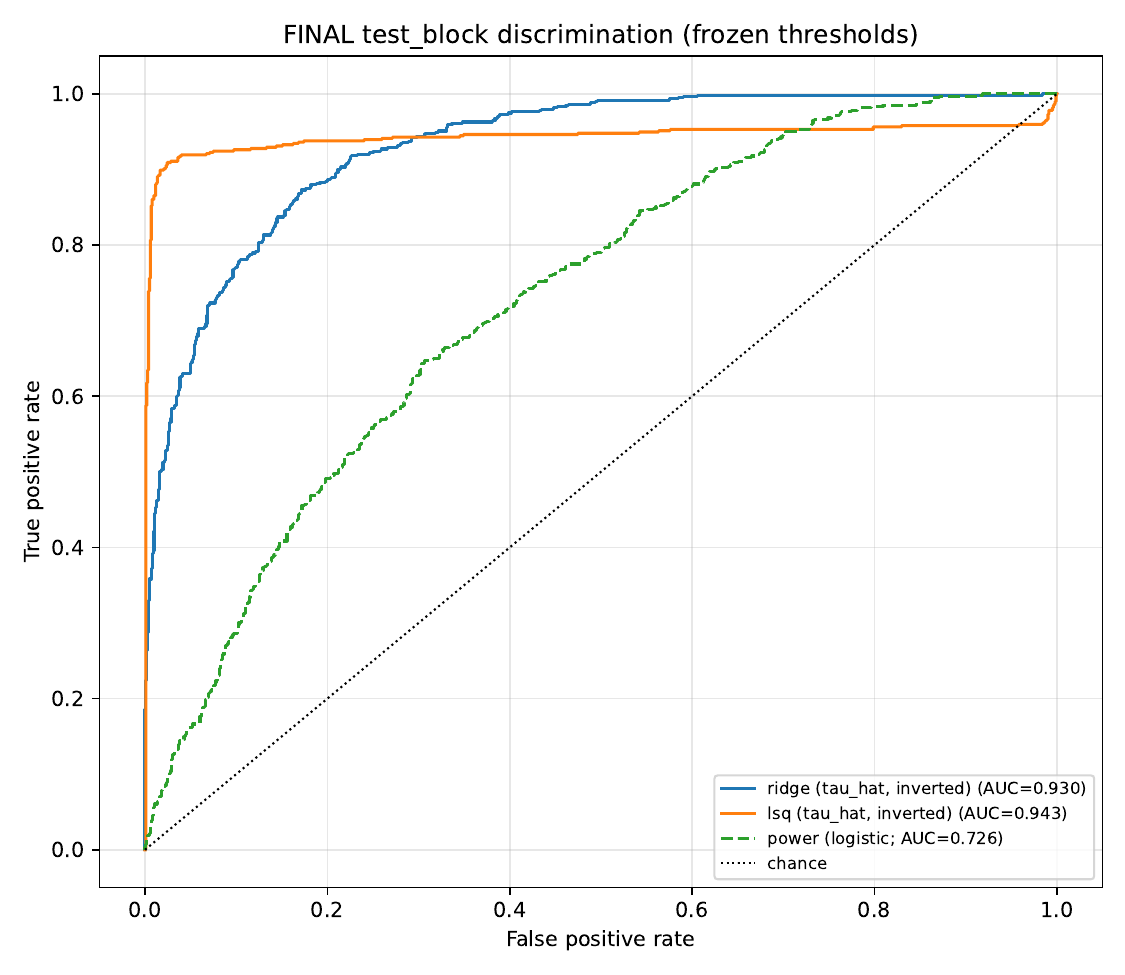}
    \caption{Final \texttt{test\_block} ROC comparison, frozen thresholds, evaluated once after the
    design freeze. LSQ (AUC$=0.943$) and ridge (AUC$=0.930$) both substantially outperform the calibrated
    power baseline (AUC$=0.726$) in ranking quality; Table~\ref{tab:final_results} shows the
    operating-point recall gap is far larger still.}
    \label{fig:final_roc}
\end{figure}

\begin{table*}[htbp]
    \centering
    \caption{Final \texttt{test\_block} evaluation (held out until design freeze; evaluated once). TPR/FPR
    are reported with Wilson 95\% confidence intervals; AUC is defined only on the Stage-1-passing
    (conditional) set. See Sec.~\ref{sec:stage2} for method definitions. Ridge/energy-decay is a
    TF-native, interpretable, secondary Stage-2 method; nonlinear LSQ is the strongest Stage-2
    discriminator (primary); the calibrated power baseline is a weaker, direction-sensitive baseline with
    AUC(power)$=0.274$, AUC($-$power)$=0.726$.}
    \label{tab:final_results}
    \begin{tabular}{@{}lllll@{}}
        \toprule
        \textbf{Method} & \textbf{Mode} & \textbf{TPR [95\% CI]} & \textbf{FPR [95\% CI]} & \textbf{AUC} \\
        \midrule
        Ridge/energy-decay & Conditional & 0.486~[0.446, 0.526] & 0.016~[0.010, 0.025] & 0.930 \\
        Ridge/energy-decay & End-to-end & 0.482~[0.442, 0.522] & 0.013~[0.008, 0.022] & -- \\
        Nonlinear LSQ & Conditional & 0.879~[0.850, 0.903] & 0.012~[0.007, 0.020] & 0.943 \\
        Nonlinear LSQ & End-to-end & 0.872~[0.843, 0.896] & 0.010~[0.006, 0.017] & -- \\
        Power (calibrated logistic) & Conditional & 0.054~[0.038, 0.075] & 0.010~[0.005, 0.018] & 0.726 \\
        Power (calibrated logistic) & End-to-end & 0.053~[0.038, 0.074] & 0.008~[0.005, 0.015] & -- \\
        \bottomrule
    \end{tabular}
\end{table*}

Table~\ref{tab:per_tau} breaks these results down by damping time. The Stage-1 pass rate itself is
strongly, and asymmetrically, tau-dependent: among standard-damping candidates, the pass rate falls
monotonically from $97.3\%$ at $\tau=5$~ms to $66.3\%$ at $\tau=40$~ms, while both anomalous damping
times pass at $\approx99\%$. This is consistent with, and plausibly explained by, the graph statistic's
sensitivity to spatially compact time-frequency structure (Sec.~\ref{sec:stage1}): a longer-duration
signal spreads the same matched-filter-normalized energy across more time frames, lowering the local
brightness the statistic is built to reward. We regard this as both a limitation --- roughly a third of
genuine long-duration standard signals are rejected before Stage~2 ever evaluates them, which affects the
end-to-end numbers in Table~\ref{tab:final_results} --- and a favorable, unengineered property for this
application, since it happens to preferentially retain exactly the short-duration signals this screening
task is designed to flag. At the level of Stage-2 discrimination itself, the least-squares estimator's
advantage over the ridge estimator is concentrated specifically at the harder $\tau=3$~ms case (flag rate
$0.826$ versus $0.237$), while both perform comparably at the easier $\tau=1$~ms case ($0.932$ versus
$0.736$); the least-squares estimator's false-flag rate on standard-damping candidates is exactly zero
for $\tau\in\{10,20,40\}$~ms.

\begin{table*}[htbp]
    \centering
    \caption{Per-$\tau$ breakdown on \texttt{test\_block}: Stage-1 pass rate and Stage-2 conditional flag
    rate for each method. For anomalous $\tau$, the flag rate is a recall (higher is better); for standard
    $\tau$, it is a false-flag rate (lower is better).}
    \label{tab:per_tau}
    \begin{tabular}{@{}llllll@{}}
        \toprule
        $\boldsymbol{\tau}$ \textbf{(ms)} & \textbf{Class} & \textbf{Stage-1 pass rate} & \textbf{LSQ flag rate} & \textbf{Ridge flag rate} & \textbf{Power flag rate} \\
        \midrule
        1  & Anomalous & 0.987~[0.966, 0.995] & 0.932~[0.898, 0.956] & 0.736~[0.684, 0.783] & 0.054~[0.034, 0.086] \\
        3  & Anomalous & 0.997~[0.981, 0.999] & 0.826~[0.779, 0.865] & 0.237~[0.193, 0.289] & 0.054~[0.033, 0.085] \\
        5  & Standard  & 0.973~[0.948, 0.986] & 0.041~[0.024, 0.070] & 0.024~[0.012, 0.049] & 0.021~[0.009, 0.044] \\
        10 & Standard  & 0.937~[0.903, 0.959] & 0.000~[0.000, 0.013] & 0.014~[0.006, 0.036] & 0.011~[0.004, 0.031] \\
        20 & Standard  & 0.853~[0.809, 0.889] & 0.000~[0.000, 0.015] & 0.008~[0.002, 0.028] & 0.004~[0.001, 0.022] \\
        40 & Standard  & 0.663~[0.608, 0.714] & 0.000~[0.000, 0.019] & 0.015~[0.005, 0.043] & 0.000~[0.000, 0.019] \\
        \bottomrule
    \end{tabular}
\end{table*}

Held-out evaluation on the final segment was consistent with, and on most metrics marginally better than,
the earlier within-training-data validation split (Sec.~\ref{sec:final_eval}); no threshold or parameter
was adjusted after this evaluation.

\section{Limitations}
\label{sec:limitations}

\begin{figure*}[htbp]
    \centering
    \includegraphics[width=0.85\textwidth]{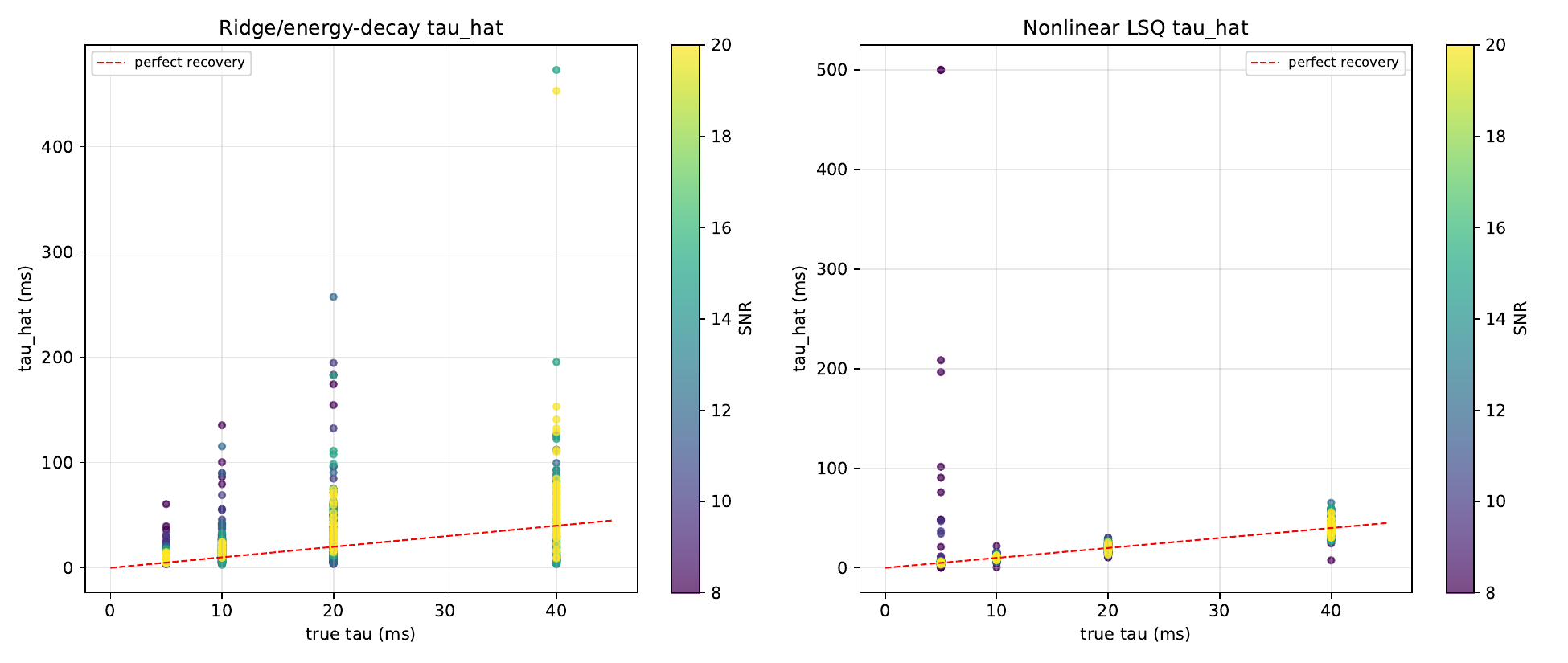}
    \caption{Recovered $\hat\tau$ versus true $\tau$ on standard-damping candidates, ridge (left) and LSQ
    (right), colored by SNR, from a training-data validation pass. Both estimators show substantial
    scatter above the perfect-recovery line, especially at low SNR, supporting this section's point that
    $\hat\tau$ is validated as an anomaly-ranking score (Sec.~\ref{sec:results}), not as a precise point
    estimate of the true damping time.}
    \label{fig:tau_recovery}
\end{figure*}

\begin{itemize}
    \item \textbf{Signal model.} The injected waveform (Eq.~\ref{eq:signal_model}) is a single-mode
    phenomenological proxy, not a physically complete post-merger waveform; it does not capture
    multi-mode structure, frequency drift, or amplitude modulation present in numerical-relativity
    post-merger signals \cite{Bauswein2016HighDensityMatter,Whittaker2022PostMergerML}. A particularly
    important omitted effect is frequency evolution. If a real post-merger signal drifts in frequency
    while the Stage-2 least-squares estimator assumes a constant frequency, accumulated phase mismatch
    could be absorbed by the fit as apparent amplitude decay, biasing the recovered effective damping
    time toward shorter values -- the same direction as the anomaly this pipeline is designed to flag.
    The present work therefore does not establish robustness of the Stage-2 damping score to realistic
    frequency-drifting post-merger waveforms; testing this directly, with chirped and
    numerical-relativity-informed injections, is future work (Sec.~\ref{sec:conclusion}).
    \item \textbf{Single detector.} All results use LIGO Livingston (L1) data only; no multi-detector
    coincidence or network-level significance is computed or claimed.
    \item \textbf{Stage-1 interpretation.} The Stage-1 statistic is validated as sensitive to local
    time-frequency organization beyond total power (Sec.~\ref{sec:stage1}); it is not validated as a
    measure of physical post-merger coherence specifically, and the shuffle test shows real detector
    noise itself already has non-trivial local structure, so any physical interpretation must remain
    comparative rather than absolute.
    \item \textbf{$\hat\tau$ is a ranking score, not a validated point estimate.} As Fig.~\ref{fig:tau_recovery}
    shows directly, both Stage-2 estimators exhibit substantial scatter above the perfect-recovery line,
    especially at low SNR. The evidence in this work supports $\hat\tau$ as an anomaly-ranking score
    (Sec.~\ref{sec:results}) --- it discriminates standard from anomalous damping effectively --- but does
    not support treating a single $\hat\tau$ value as an accurate point estimate of the true physical
    damping time.
    \item \textbf{Reported false-positive rates are finite-sample, not production search rates.} The
    FPR values in Tables~\ref{tab:final_results} and~\ref{tab:per_tau} are finite-sample false-flag rates
    on constructed injection/noise windows, not production gravitational-wave search false-alarm rates
    estimated over months or years of detector background. We do not claim production search sensitivity
    or coherent-WaveBurst/PyCBC-level false-alarm-rate calibration.
    \item \textbf{Calibration sample sizes.} Several calibrated quantities (in particular the Stage-1
    null threshold and both Stage-2 anomaly thresholds) rely on percentile estimates whose precision is
    limited by calibration sample size; we report this explicitly rather than presenting point thresholds
    as exact.
    \item \textbf{No learned-classifier baseline.} Neither a simple convolutional neural network nor the
    frozen-audio-transformer approach explored in an earlier version of this project is included as a
    baseline here; both are out of scope for this version and would require their own independently fair
    training/validation/held-out-test protocol if added in future work.
    \item \textbf{No dark matter or dark-sector detection claim.} This study establishes a statistically
    defensible screening methodology for a phenomenological damping-time anomaly on synthetic, injected
    signals; it does not claim, and should not be read as claiming, evidence for dark matter, dark-sector
    physics, or any specific non-standard energy-loss mechanism.
\end{itemize}

\section{Conclusion}
\label{sec:conclusion}

We have presented a two-stage, detectability-aware screening framework for anomalous post-merger damping
signatures, validated end-to-end on real LIGO L1 noise with an empirically calibrated null distribution,
a synthetic-Gaussian-noise formula check, a structure-versus-power ablation, and a held-out final
evaluation performed only after the full design was frozen. The central, conservatively stated result is
two-fold and deliberately separates the two stages' roles: the graph-spectral Stage-1 statistic
substantially outperforms total power as a detectability gate for compact, short-damping injections.
Conditional on this gate, damping-time-based Stage-2 scores, especially the nonlinear least-squares
estimator, distinguish anomalously short damping proxies from the standard damping manifold far more
effectively than a calibrated power-only baseline. Future work includes extending the signal model
toward numerical-relativity-informed post-merger waveforms, multi-detector coincidence, and a fair,
separately validated comparison against learned time-frequency classifiers.

\appendix

\section{Reproducibility Details}
\label{app:reproducibility}

The values below are transcribed directly from \texttt{frozen\_config.json}, the machine-readable file
governing every threshold, model, and hyperparameter used in the final evaluation (Sec.~\ref{sec:final_eval}
and Sec.~\ref{sec:results}); that file, released alongside this manuscript's code repository
(Sec.~\ref{app:data_code}), is authoritative if it and this appendix ever disagree. All data-processing,
calibration, and evaluation notebooks referenced throughout this work are released in the same
repository; each is self-contained and documents its own dependencies on earlier stages.

\subsection{Stage 1}
Score: $\mathrm{norm\_lambda1}$ (Sec.~\ref{sec:stage1}). Threshold: $55.6439$, at $\alpha=0.01$,
calibrated on \texttt{calib\_block}. This threshold value was confirmed identical across two
independent re-derivations of the calibration procedure (once during Stage-2 development, once during
the final evaluation), which is why Figs.~\ref{fig:null_distributions} and~\ref{fig:phase_diagram} ---
generated from an earlier, smaller-sample calibration pass showing $54.98$ --- are noted there as
exploratory rather than the exact frozen value.

\subsection{Stage 2, primary --- nonlinear LSQ}
Nonlinear damped-sinusoid least-squares fit (\texttt{scipy.optimize.curve\_fit}) on the whitened
time-domain window. Anomaly threshold: $\hat\tau < 3.472$~ms. Fit/calibration pool: the Stage-2
calibration half of the training segment (\texttt{stage2\_calib}, Table~\ref{tab:data_split}).

\subsection{Stage 2, secondary --- ridge/energy-decay}
Anomaly threshold: $\hat\tau < 4.365$~ms, fit on the same \texttt{stage2\_calib} pool. Estimator
parameters (frozen as of the held-out validation stage, Sec.~\ref{sec:final_eval}; not tuned
thereafter): maximum fit window $60$ frames; minimum frames required for a fit, $4$; noise-floor
``patience'' (consecutive below-floor frames required to truncate the fit), $2$; noise floor set at the
$90$th percentile of \texttt{calib\_block} frame-power samples, value $274565.08$ (arbitrary CQT power
units, Sec.~\ref{sec:stage1}).

\subsection{Power baseline}
Single-feature logistic regression on $\mathrm{total\_power\_raw}$ (in-band CQT power,
$1500$--$4000$~Hz), fit on \texttt{stage2\_calib}. Operating threshold: predicted
$P(\mathrm{anomalous}) > 0.3656$. Regression coefficients are not persisted as raw numbers in the frozen
configuration; they are re-derived deterministically from \texttt{stage2\_calib} in the evaluation
notebook, and the resulting operating threshold is asserted to match the value above before any
evaluation proceeds (Sec.~\ref{sec:final_eval}).

\subsection{Time-frequency representation and graph construction}
CQT: $f_{\rm min}=500$~Hz, $48$ bins, $12$ bins per octave, hop length $8$ samples, restricted to the
$1500$--$4000$~Hz band for the Stage-1 statistic (Sec.~\ref{sec:stage1}; cf.\ Fig.~\ref{fig:cqt_examples}'s
wider visual range). Graph: each time-frequency tile is connected to others within a local neighborhood
of radius $2$ bins in frequency and $2$ frames in time.

\subsection{Whitening}
Frequency-domain whitening band-limited to $[20, 6000]$~Hz; sample rate $16384$~Hz; analysis window
duration $100$~ms; edge trim $4.0$~s (Sec.~\ref{sec:data_psd}).

\subsection{Evaluation grid}
$\tau \in \{1,3,5,10,20,40\}$~ms, with $\{5,10,20,40\}$~ms designated standard and $\{1,3\}$~ms
designated anomalous; matched-filter SNR $\in \{8,10,12,15,20\}$.

\subsection{Explicitly excluded from this version}
A simple convolutional neural network and the frozen-audio-transformer (AST/Mahalanobis) approach from
an earlier version of this project are both excluded from the frozen design (Sec.~\ref{sec:limitations}).

\subsection{Non-load-bearing diagnostic}
The band-leakage diagnostic (raw/full-passband power ratio across $\tau$) was inconclusive and is
recorded in the frozen configuration explicitly as not to be used to justify, adjust, or interpret any
threshold above.

\subsection{Governing rules}
Recorded verbatim in \texttt{frozen\_config.json} and enforced in the final evaluation notebook: (i)
\texttt{test\_block} may only be loaded in the final evaluation notebook; (ii) no threshold, model,
hyperparameter, or preprocessing choice may be changed based on \texttt{test\_block} results; (iii) no
further tuning of the ridge estimator is permitted; (iv) if \texttt{test\_block} performance is worse
than the held-out validation split, that is reported as-is, not tuned away.

\section*{Data and Code Availability}
\label{app:data_code}

The data, code, and computational environment configurations used in this study, including the frozen
configuration file, all analysis notebooks, fixed random seeds, and instructions for reproducing every
figure and table in this manuscript, are hosted on GitHub at
\url{https://github.com/ruslanalas/gw-postmerger-detectability}. % TODO: confirm/replace repository URL
To guarantee long-term access, a tagged release of the repository is archived on Zenodo under DOI:
\href{https://doi.org/10.5281/zenodo.21993621}{10.5281/zenodo.21993621}.
Public strain data from the LIGO Livingston detector were fetched via the Gravitational Wave Open
Science Center.

\begin{acknowledgments}
% TODO: update acknowledgments for the current version of this work.
This research has made use of data, software, and/or web tools obtained from the Gravitational Wave Open
Science Center (\url{https://www.gw-openscience.org}), a service of LIGO Laboratory, the LIGO Scientific
Collaboration, and the Virgo Collaboration.
\end{acknowledgments}

\end{document}